\documentclass[%
 aip,
 amsmath,amssymb,
reprint, twocolumn 
]{revtex4-1}

\usepackage[dvipdfmx]{graphicx}
\usepackage{dcolumn}
\usepackage{bm}

\usepackage[utf8]{inputenc}
\usepackage[T1]{fontenc}
\usepackage{mathptmx}
\usepackage{etoolbox}
\usepackage{color}
\usepackage{ulem}
\newcommand{\sgn}{\mathop{\mathrm{sgn}}\nolimits}

\makeatletter
\def\@email#1#2{%
 \endgroup
 \patchcmd{\titleblock@produce}
  {\frontmatter@RRAPformat}
  {\frontmatter@RRAPformat{\produce@RRAP{*#1\href{mailto:#2}{#2}}}\frontmatter@RRAPformat}
  {}{}
}%
\makeatother
\begin{document}

\preprint{AIP/123-QED}

\title[title]{Electric field-induced nonreciprocal spin current due to chiral phonons in chiral-structure superconductors}

\author{Dapeng Yao}
\affiliation{Department of Physics, Tokyo Institute of Technology, 2-12-1 Ookayama, Meguro-ku, Tokyo 152-8551, Japan}

\author{Mamoru Matsuo}
\affiliation{Kavili Institute for Theoretical Sciences, University of Chinese Academy of Sciences, Beijing, 100190, China}
\affiliation{CAS Center for Excellence in Topological Quantum Computation, University of Chinese Academy of Sciences, Beijing, 100190, China}
\affiliation{Advanced Science Research Center, Japan Atomic Energy Agency, Tokai, 319-1195, Japan}
\affiliation{RIKEN Center for Emergent Matter Science (CEMS), Wako, Saitama 351-0198, Japan}

\author{Takehito Yokoyama}
\affiliation{Department of Physics, Tokyo Institute of Technology, 2-12-1 Ookayama, Meguro-ku, Tokyo 152-8551, Japan}

\date{\today}

\begin{abstract}
The recent experiment~[R.~Nakajima, $et$ $al$., Nature \textbf{613}, 479 (2023)]~has reported a pair of oppositely polarized spins under an alternating  electric current in a superconductor with a chiral structure. However, these behaviors cannot be explained by the conventional Edelstein effect and require a new mechanism.
In this Letter, we propose a mechanism of spin current generation under an external electric field due to chiral phonons in a chiral-structure superconductor based on the Bogoliubov de Gennes and the Boltzmann equations.
In our mechanism, chiral phonons are induced by electric field due to inversion symmetry breaking and electron-phonon interaction.
They work as an effective Zeeman field and hence spin-polarize Bogoliubov quasiparticles in the superconductor.
As a result, the spin current carried by quasiparticles flows along the screw axis and shows a quadratic dependence on the electric field at the low-field range, leading to a nonreciprocal spin transport.
The spin current also shows a nonmonotonic temperature dependence and has a maximum at around the superconducting transition temperature.
\end{abstract}

\maketitle


Recently, the studies on chiral phonons have attracted much attention. Since the prediction of a circularly polarized phonon mode with angular momentum due to lattice symmetry constrains~\cite{LZhang2014,LZhang2015,Chen2019,Rostami2022}, chiral phonons have been observed~\cite{HZhu2018,Ishito2023,Grissonnanche2020,XTChen2019,Li2019,Ueda2023,Jeong2022} and their various aspects have been revealed.
Phonon angular momentum can be generated by temperature gradient when inversion symmetry is broken, dubbed as phonon thermal Edelstein effect~\cite{Hamada2018}, and also by an electric field via the lattice distortion~\cite{Hamada2020}.
Furthermore, since ions have the charges, chiral phonons accompanying atomic rotations in crystals act as an effective magnetic field.~\cite{Nova2017,Juraschek2019,Geilhufe2021,Juraschek2022,Xiong2022,Luo2023,Hernandez2022,Chaudhary2023}
The induced magnetic fields can be large, e.g., on the order of one tesla in cerium fluoride~\cite{Luo2023}.
Coupling between chiral phonons, electrons, magnons or spins can induce or change spin magnetization~\cite{HamadaPRR2020,Ren2021,Yao2022,Tauchert2022,Yao2023,Fransson2023,Davies2023,YafenRen2023,Kahana2023}.
The recent experiment has observed a chiral-phonon-activated spin Seebeck effect: spin current generated by a temperature gradient in a non-magnetic chiral material~\cite{Kim2023}. Here, spontaneous magnetization is induced by chiral phonons, leading to spin current generation.
Also, theoretically, the spin Seebeck effect due to non-equilibrium distribution of chiral phonons under a temperature gradient in chiral materials has been proposed~\cite{Li2022}.

In addition to non-magnetic chiral materials in the normal states, a recent experiment has reported a pair of oppositely polarized spins in an electric alternating current (a.c) in a superconductor with a chiral structure~\cite{Nakajima2023}. However, the conventional Edelstein effect cannot consistently explain this behavior~\cite{Edelstein1990,Edelstein1995,Edelstein2005,He}: the spin accumulation at the two ends of the superconductor with the antiparallel spins. Thus, this requires a new mechanism of spin current generation.

In this Letter, we theoretically propose a generation mechanism of spin current due to chiral phonons under an external electric field in a chiral-structure superconductor based on the Bogoliubov de Gennes (BdG) and the Boltzmann transport equations.
In our mechanism, due to inversion symmetry breaking and electron-phonon interaction, chiral phonons are generated by an electric field and play a role of effective Zeeman magnetic field, which spin-polarizes quasiparticles. We find that a spin current carried by the Bogoliubov quasiparticles flows along the screw axis under the electric field and shows a quadratic dependence on the electric field at the low-field range, resulting in a nonreciprocal spin transport. 
The spin current also shows a nonmonotonic temperature dependence and has a maximum at around the superconducting transition temperature.
These behaviors are qualitatively consistent with the recent experiment.\cite{Nakajima2023}

\begin{figure}[htb]
\begin{center}
\includegraphics[clip,width=8.5cm]{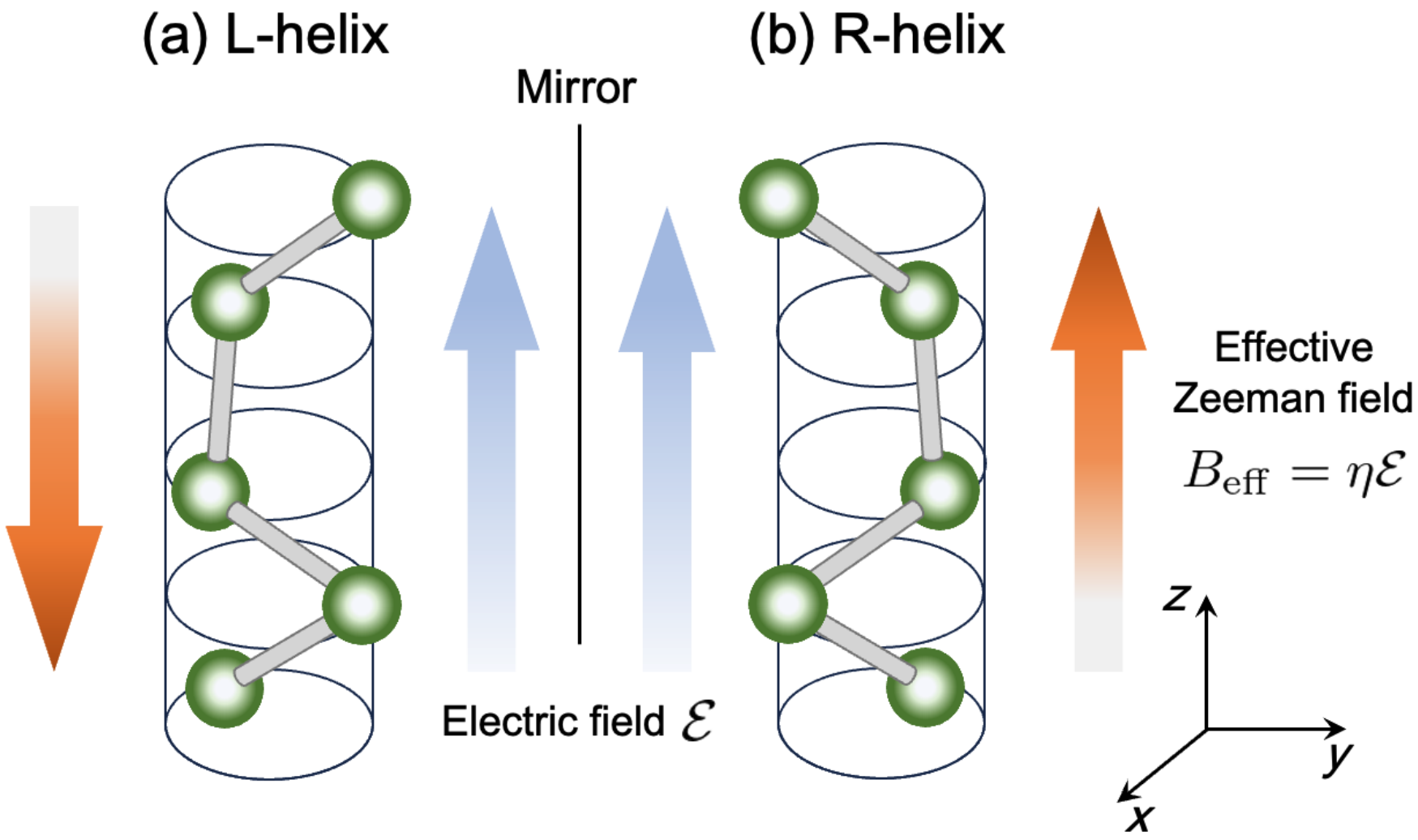}
\end{center}
\caption{
(Color online). Superconductor with a chiral structure under an electric field $\mathcal{E}$. An effective Zeeman field $B_{\rm{eff}}=\eta\mathcal{E}$ is generated along the screw axis, and the directions of $B_{\rm{eff}}$ are opposite for the L(left)-helix and R(right)-helix structure.}
\label{struct}
\end{figure}

Here, we consider an $s$-wave bulk superconductor with left-handed or right-handed crystal structures under an external electric field as shown in Fig.~\ref{struct}.
Under the electric field, quasiparticles flow. With the electron-phonon interaction, phonons are dragged~\cite{Perrin1974,Gurevich1989}, and their non-equilibrium distributions make the phonon angular momentum nonzero~\cite{Hamada2018}. Thus, the ions rotate around their equilibrium positions, carrying a circular current, which results in an effective magnetic field~\cite{Nova2017,Juraschek2019,Geilhufe2021,Juraschek2022,Xiong2022,Luo2023,Hernandez2022,Chaudhary2023}.
This effective magnetic field splits the electronic band structures via the Zeeman coupling. 
We suppose that the external electric field $\mathcal{E}$ is applied along the screw axis, and then the effective Zeeman field $B_{\rm{eff}}$ is generated.
The effective Zeeman field is parallel or antiparallel to the electric field
with a phenomenological constant $\eta$: $B_{\rm{eff}}=\eta\mathcal{E}$. Here, the sign of $\eta$ is determined by the chirality of the crystal, i.e.,  opposite for left-handed and right-handed crystals.
Nonzero $\eta$ is allowed for gyrotropic crystals.
Among 21 point groups lacking inversion symmetry, 18 point groups are gyrotropic (from which 11 point groups are chiral)\cite{Lozel1975,Jerphagnon1976,Ganichev2016}.

Since $\eta$ stems from the coupling between chiral phonons and electrons, it should depend on the temperature. We assume that it decreases with increasing temperature due to the enlarged distance between the electrons and ions because vibrational motions of ions become larger when the temperature increases~\cite{Gao2023}. We adopt the temperature dependence of $\eta$ given in Ref.~\cite{Gao2023} as follows:
\begin{align}
\eta=\eta_0e^{-T/T_c}
\end{align}
with a constant $\eta_0$.

In terms of the BdG formulation, the Hamiltonian of the superconductor can be written as $H=\frac{1}{2}\sum_{\bm{k}}\Psi^{\dagger}_{\bm{k}}\mathcal{H}(\bm{k})\Psi_{\bm{k}}$ with the operator $\Psi^{\dagger}_{\bm{k}}=(c^{\dagger}_{\bm{k},\uparrow},c^{\dagger}_{\bm{k},\downarrow},c_{-\bm{k},\uparrow},c_{-\bm{k},\downarrow})$, where $c^{\dagger}_{\bm{k},\sigma}~(c_{\bm{k},\sigma})$ represents the creation (annihilation) operator of the electron with spin $\sigma$. 
Under the external electric field $\mathcal{E}$, the Zeeman energy-splitting term as we mentioned above for each spin is included in the BdG Hamiltonian
\begin{align}
\mathcal{H}(\bm{k})=
\begin{pmatrix}
\xi_{\bm{k}}+\eta\mathcal{E} & 0 & 0 & \Delta \\
0 & \xi_{\bm{k}}-\eta\mathcal{E} & -\Delta & 0 \\
0 & -\Delta & -\xi_{-\bm{k}}-\eta\mathcal{E} & 0 \\
\Delta & 0 & 0 & -\xi_{-\bm{k}}+\eta\mathcal{E}
\end{pmatrix},
\end{align}
where $\Delta$ is the superconducting gap function, and $\xi_{\bm{k}}=\frac{\hbar^2k^2}{2m}-\mu$ with $\mu$ denoting the chemical potential. It gives the energy eigenvalues as
\begin{align}
\label{energy}
E_{\pm\bm{k},\sigma}=\pm\sqrt{\xi_{\bm{k}}^2+\Delta^2}+\sigma\eta\mathcal{E}.
\end{align}

Here, in the presence of the electric field $\mathcal{E}$, the distribution function of Bogoliubov quasiparticles $f_{\bm k\sigma}$ changes into $f_{\bm k\sigma}=f_{\bm k\sigma}^0+\delta f_{\bm k\sigma}$, where $f_{\bm k\sigma}^0=f(E_{\bm k\sigma})=(e^{\beta E_{\bm k\sigma}}+1)^{-1}$ is the Fermi distribution function in equilibrium. By means of the Boltzmann equation and relaxation time approximation, the change of the distribution function can be written as
\begin{align}
\delta f_{\bm k\sigma}=\sgn(\xi_{\bm k})e\mathcal{E}v_{\bm k\sigma}\tau_{\rm{sc}}\frac{\partial f_{\bm k\sigma}}{\partial E_{\bm k\sigma}}\simeq \sgn(\xi_{\bm k})e\mathcal{E}v_{\bm k\sigma}\tau_{\rm{sc}}\frac{\partial f_{\bm k\sigma}^0}{\partial E_{\bm k\sigma}},
\end{align}
where $\sgn(\xi_{\bm k})$ appears because electrons and holes have the opposite sign of charge, and $\tau_{\rm{sc}}$ represents the relaxation time in the superconducting state and can be related to the relaxation time in the normal state $\tau_{\rm{n}}$ as $\tau_{\rm{sc}}=\left(\sqrt{\xi_{\bm k}^2+\Delta^2}/|\xi_{\bm k}|\right)\tau_{\rm{n}}$~\cite{Tinkham,Yamashita2002}. Here, the velocity of Bogoliubov quasiparticle in the superconductor is
\begin{align}
v_{\bm k\sigma}=v_{\bm k}=\frac{1}{\hbar}\frac{\partial E_{\bm k\sigma}}{\partial k}=\frac{\xi_{\bm k}}{\sqrt{\xi_{\bm k}^2+\Delta^2}}\frac{\hbar k}{m},
\end{align}
which is independent of the electron spin. Because of the changes of the distribution functions for up-spin and down-spin Bogoliubov quasiparticles under the electric field, the spin current carried by Bogoliubov quasiparticles is naturally generated in the bulk superconductor. The spin current can be expressed as
\begin{align}
\label{sc}
j_s=&e\int\frac{d^3k}{(2\pi)^3}v_{\bm k}\left(\delta f_{\bm k\uparrow}-\delta f_{\bm k\downarrow}\right) \nonumber \\
=&e^2\tau_{\rm{n}}\mathcal{E}\int\frac{d^3k}{(2\pi)^3}\frac{\sqrt{\xi_{\bm{k}}^2+\Delta^2}}{\xi_{\bm{k}}}v^2_{\bm{k}}\left(\frac{\partial f^0_{\bm{k},\uparrow}}{\partial E_{\bm{k},\uparrow}}-\frac{\partial f^0_{\bm{k},\downarrow}}{\partial E_{\bm{k},\downarrow}}\right)
\end{align} 
\begin{widetext}
\begin{eqnarray}
\label{sc_t}
=-\frac{e^2\tau_{\rm{n}}\sqrt{2m}\mathcal{E}}{4\pi^2\hbar^3k_BT}\int_{-\mu}^{\xi_c}d\xi\frac{\xi}{\sqrt{\xi^2+\Delta^2}}(\xi+\mu)^{\frac{3}{2}}
\Bigg\{\cosh^{-2}\left(\frac{\sqrt{\xi^2+\Delta^2}+\eta\mathcal{E}}{2k_BT}\right)-\cosh^{-2}\left(\frac{\sqrt{\xi^2+\Delta^2}-\eta\mathcal{E}}{2k_BT}\right)\Bigg\},
\end{eqnarray}
\end{widetext}
with the cufoff $\xi_c$. Some details on this calculation are given in Appendix.  
Here, we do not consider contribution from Cooper pairs to the spin current since singlet Cooper pairs have zero spin angular momentum.

In the numerical calculation, we set the cutoff energy $\xi_c=0.45$eV, the chemical potential $\mu=1.28$eV, the relaxation time in the normal state $\tau_{\rm n}=10$ps, and the mass $m=9.11\times10^{-31}$kg.
In the recent experiment, an a.c electric field with the upper limit of $10^6$V/m has been applied to the chiral-structure superconductor with a superconducting transition temperature $T_c$ of 7.5K~\cite{Nakajima2023}.

\begin{figure}[htb]
\begin{center}
\includegraphics[clip,width=9cm]{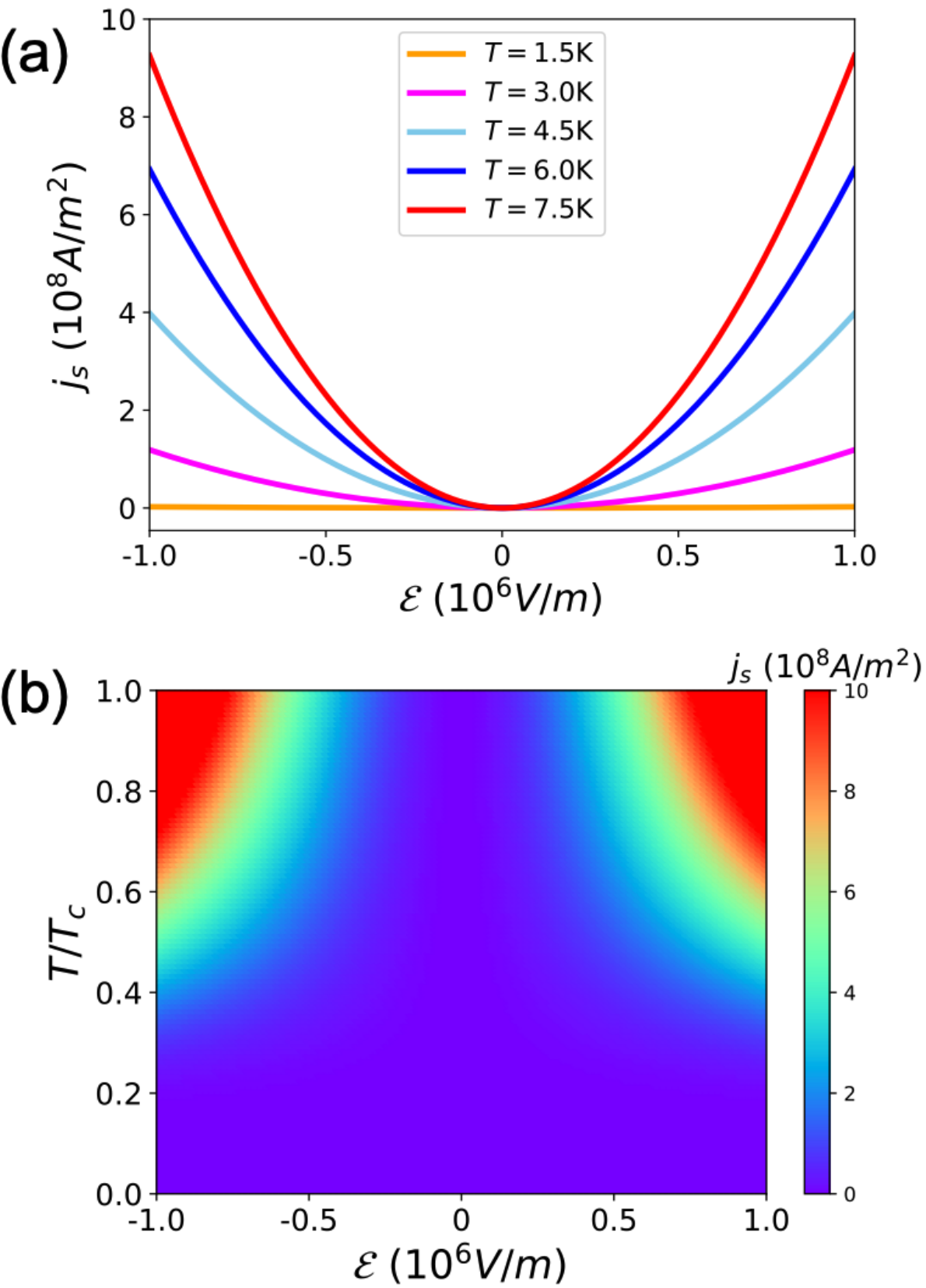}
\end{center}
\caption{
(Color online). (a) Electric field $\mathcal{E}$ dependence of spin current $j_s$. (b)  $j_s$ as functions of $\mathcal{E}$ and $T$. Here, the superconducting transition temperature $T_c$ is 7.5K and $\eta_0$ is chosen as $10^{-10}\rm{eV\cdot m/V}$.}
\label{spin_current}
\end{figure}

Figure~\ref{spin_current}(a) shows the electric field $\mathcal{E}$ dependence of the spin current $j_s$ in Eq.~(\ref{sc_t}) with different temperatures in superconducting states. The spin current as a function of electric field and temperature below the superconducting transition temperature $T_c$ is also presented in Fig.~\ref{spin_current}(b).
Here, we choose $\eta_0$  as $10^{-10}\rm{eV\cdot m/V}$. Then, the Zeeman splitting energy is $0.1$meV for $\mathcal{E}=10^{6}$V/m~\cite{Li2022} at $T=0$.
Figures~\ref{spin_current}(a)(b) show that the spin current is increased with the increase of the electric field or temperature.
We also find that the spin current $j_s$ is an even function of the electric field $\mathcal{E}$ and hence the spin transport is nonreciprocal.
The spin current $j_s$ exhibits a quadratic dependence on the electric field $\mathcal{E}$ at the low-field range.

\begin{figure}[htb]
\begin{center}
\includegraphics[clip,width=8cm]{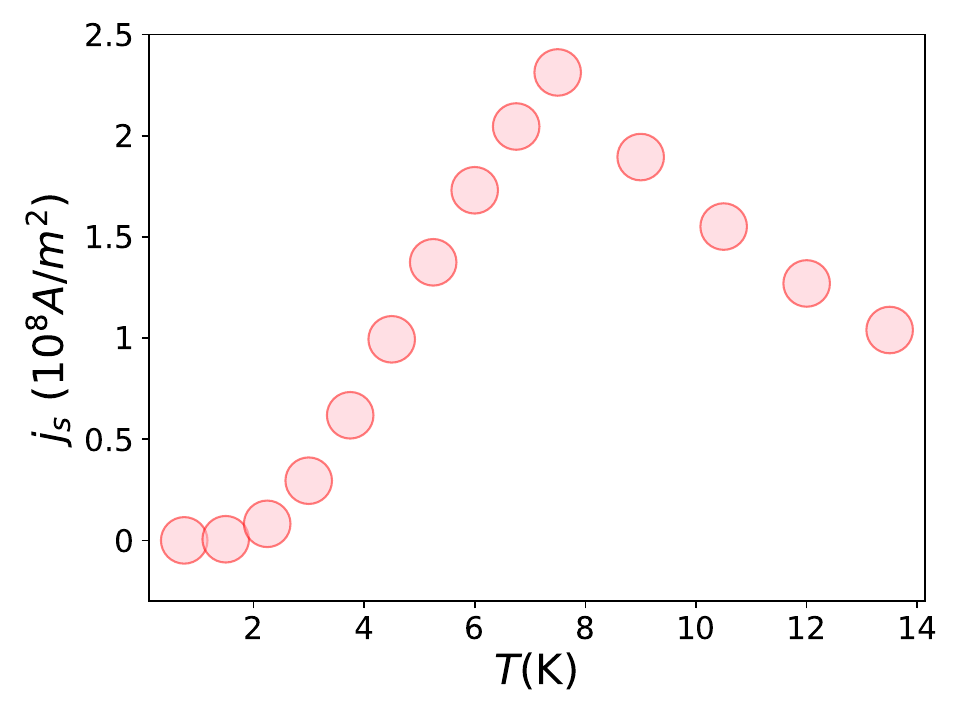}
\end{center}
\caption{
(Color online). Temperature dependence of spin current for 1.5K~$\leq T\leq$~13.5K. Here, the electric field is $0.5 \times10^6$V/m. }
\label{tem_depen}
\end{figure}

Next, we show the temperature dependence of the spin current.
The recent experiment has clarified that the anti-parallel pair of the spin polarization at the edges is maximum at $T=7.4\rm{K}$, i.e. the vicinity of the superconducting transition temperature, and it quickly decays when the temperature deviates from the superconducting transition temperature $T_c=7.5\rm{K}$~\cite{Nakajima2023}. In our model, the spin current $j_s$ shows a similar temperature dependence and becomes maximum around $T_c=7.5\rm{K}$ as shown in Fig.~\ref{tem_depen}. 
This behavior can be understood as follows.
In the superconducting states, when the temperature increases, the superconducting gap function becomes smaller, and quasiparticles are more excited. As a result, the contribution from quasiparticles to spin current becomes larger. 
When the temperature exceeds the superconducting transition temperature, the superconducting gap becomes zero. Then, the spin current is smeared out by the derivative of the Fermi distribution function in Eq.~(\ref{sc_t}) and suppressed by the reduced electron-chiral phonon coupling~\cite{Gao2023} at higher temperature (Eq.(1)), and hence the spin current decreases with increasing temperature.
In this way, we understand that the spin current takes maximum at around $T_c$ as shown in Fig.~\ref{tem_depen}.
\begin{figure}[htb]
\begin{center}
\includegraphics[clip,width=8cm]{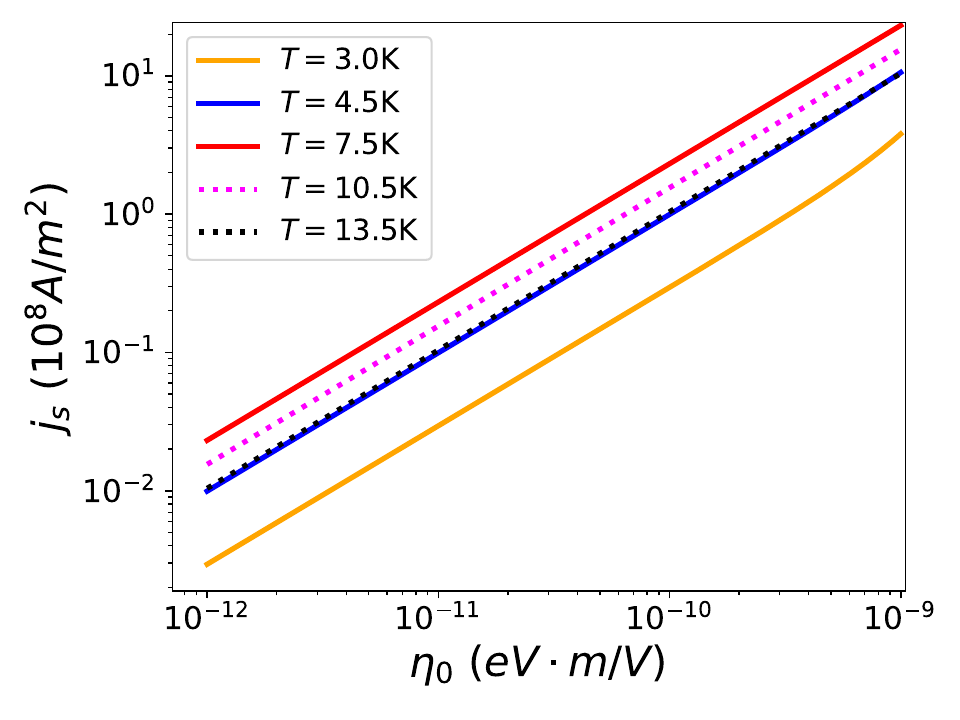}
\end{center}
\caption{
(Color online). $\eta_0$ dependence of spin current. Here, the electric field $\mathcal{E}$ is $0.5 \times10^6$V/m. The solid lines and dashed lines represent the spin current below and above $T_c$, respectively.}
\label{eta_depen}
\end{figure}

In addition to the electric field and temperature dependences measured in the experiment~\cite{Nakajima2023}, 
it is necessary to discuss dependence of the spin current on the constant $\eta_0$, which represents the strength of the effective magnetic field generated by the external electric field.
The magnitudes of the spin current $j_s$ as a function of the constant $\eta_0$ are shown in Fig.~\ref{eta_depen} by a log-log scaled plot with $\eta_0$ being $10^{-12}\sim 10^{-9}\rm{eV\cdot m/V}$ under the fixed electric field $\mathcal{E}=0.5\times 10^6\rm{V/m}$, which corresponds to the Zeeman splitting energy of the order of $1\rm{\mu eV}\sim 1\rm{meV}$. In this case, we can estimate that an effective magnetic field of $0.01\rm{T}\sim 10T$ is generated by chiral phonons in crystals under the electric field.
For comparison, let us refer to other estimations of  effective magnetic field by chiral phonons: over 100T in cerium trichloride~\cite{Juraschek2022}, around one T in cerium fluoride~\cite{Luo2023}, and about 0.01T in tellurium~\cite{Xiong2022}.

\begin{figure}[htb]
\begin{center}
\includegraphics[clip,width=6.5cm]{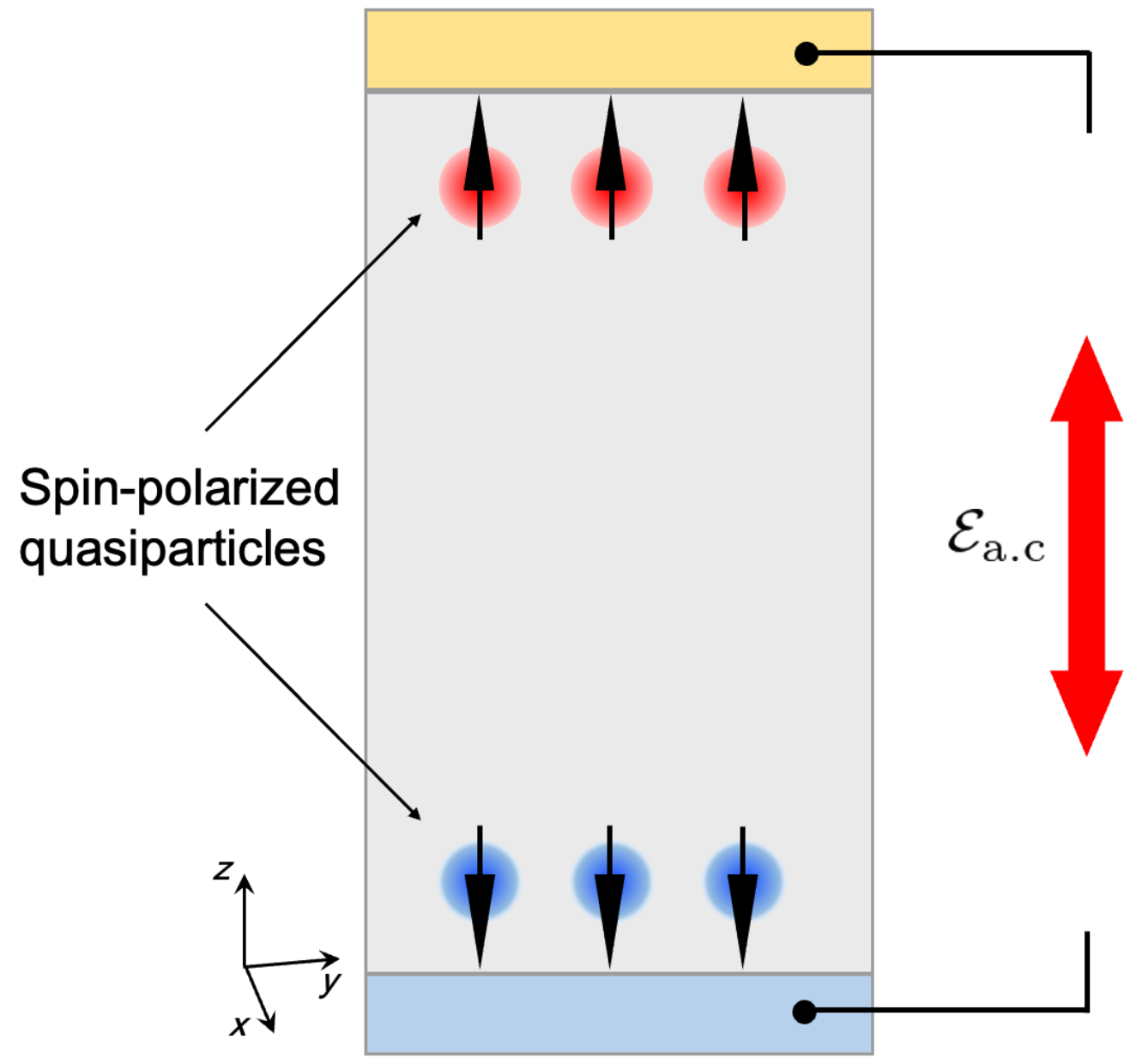}
\end{center}
\caption{
(Color online). Schematic illustration of the spin-polarized quasiparticles accumulated at the edges of the superconductor under an a.c electric field.}
\label{discussion}
\end{figure}

In our model, $\eta$ connects the effective magnetic field $B_{\rm{eff}}$ and the electric field $\mathcal{E}$.
Because $B_{\rm{eff}}$ is an axial vector and $\mathcal{E}$ is a polar vector, we have $B_{\rm{eff}}\rightarrow-B_{\rm{eff}}$ and $\mathcal{E}\rightarrow\mathcal{E}$ under the mirror reflection, and thus $\eta\rightarrow-\eta$. Therefore, the sign of $\eta$ should become opposite when the chirality of crystal is changed into that of its mirror image.
Moreover, the energy of the Bogoliubov quasiparticle in the superconductor in Eq.~(\ref{energy}) is invariant under $\sigma\rightarrow -\sigma$ and $\eta\rightarrow -\eta$, which means that the polarized quasiparticles change their direction of polarization via the mirror reflection. Thus, the left-handed and right-handed crystals show the opposite spin polarizations under the same external electric field.

Our results are also valid for a.c electric field if its frequency $\omega$ satisfies $\omega\ll\frac{\Delta}{\hbar}$. Since the magnitude of gap function is typically of order of meV, a radio wave or a low-frequency microwave are applicable. In the experiment~\cite{Nakajima2023}, the frequency $\omega=10$MHz is used, and thus we have $\hbar\omega=0.01\mu$eV$\ll\Delta$ and our results would be applicable.

There are several mechanisms of spin current generations proposed by some previous studies. The magnitudes of spin current in spin pumping in a metallic system~\cite{Ando2011}, the spin-current injection in nonlocal devices~\cite{Takahashi2008}, and the spin-vorticity coupling mechanism (coupling between spins and surface acoustic waves)\cite{Tateno2020} are, respectively, estimated as $10^{4}$ A/m$^2$, $10^{10}$ A/m$^2$, and $10^{11}$ A/m$^2$.
Our model calculation shows spin current of the order of $10^{6}-10^{9}$ A/m$^2$ as shown in Fig.~\ref{eta_depen}.
Thus, the spin current by our mechanism can be detected by the present experimental techniques.
On the other hand, the spin current deriving from the coupling between the electrons and chiral phonons is essentially different from the spin polarization generated by the Edelstein effect, although both mechanisms require breaking of inversion symmetry~\cite{Edelstein1990}. The Edelstein effect stems from a spin-momentum locking associated with helicity of electrons while our proposed mechanism considers a spin-lattice locking due to the chirality of the crystal structure. This indicates that the electron-lattice interaction by chiral phonons plays an essential role in spin transport in chiral crystals.

In summary, we have theoretically proposed a generation mechanism of nonreciprocal spin current due to chiral phonons under an external electric field in a chiral-structure superconductor based on the Bogoliubov de Gennes and the Boltzmann equations. The spin current shows a quadratic dependence on electric field at the low-field range and hence the spin transport is nonreciprocal.
Therefore, under an a.c electric field, spin-polarized quasiparticles are accumulated at the edges with opposite spin polarizations as shown in Fig.~\ref{discussion}, qualitatively consistent with the recent experiment\cite{Nakajima2023}.
Effective magnetic field induced by chiral phonons has a similarity to exchange field in ferromagnets.
Therefore, our results pave the way for application of electron-phonon coupling in chiral structures to superconducting spintronics or Josephoson junctions.


\begin{acknowledgments}
D.Y. was supported by JSPS KAKENHI Grant No.~JP23KJ0926.
M.M. was supported by JSPS KAKENHI Grants No.~JP21H01800, No.~JP21H04565, and No.~JP23H01839.
T.Y. was supported by JSPS KAKENHI Grant No.~JP30578216 and the JSPS-EPSRC Core-to-Core program ``Oxide Superspin".
\end{acknowledgments}

\section*{Data Availability}
The data that support the findings of this study are available from the corresponding author upon reasonable request.

\appendix

\section{Detailed calculation of spin current}
In this Appendix, we provide more details on the calculation of spin current defined in Eq.~(\ref{sc}). The derivative of the Fermi distribution function at the finite temperature can be calculated as
\begin{align}
\label{fd_t}
\frac{\partial f^0_{\bm k\sigma}}{\partial E_{\bm k\sigma}}=\frac{-1}{k_BT}\cosh^{-2}\left(\frac{E_{\bm k\sigma}}{2k_BT}\right).
\end{align}
Here, the integral over wave vector $\bm k$ in Eq.~(\ref{sc}) can be expressed by using the spherical coordinates:
\begin{align}
\int d^3k=\int_0^{2\pi}d\phi\int_0^{\pi}d\theta\sin\theta\int_0^{k_c}dkk^2=4\pi\int_0^{k_c}dkk^2,
\end{align}
where $k_c$ represents the cut-off wave number. Then we introduce $\xi_{\bm k}=\frac{\hbar^2k^2}{2m}-\mu$, which gives
\begin{align}
\label{replace}
k^4dk=\frac{m}{\hbar^2}\left[\frac{2m}{\hbar^2}(\xi_{\bm k}+\mu)\right]^{\frac{3}{2}}d\xi_{\bm k}.
\end{align}
The integral over the wave number $k$ can be replaced by the integral over $\xi_{\bm k}$. By substituting Eqs.~(\ref{fd_t}) and (\ref{replace}) into Eq.~(\ref{sc}), the spin current can be finally expressed by Eq.~(\ref{sc_t}).


\begin{thebibliography}{42}
\expandafter\ifx\csname natexlab\endcsname\relax\def\natexlab#1{#1}\fi
\expandafter\ifx\csname bibnamefont\endcsname\relax
  \def\bibnamefont#1{#1}\fi
\expandafter\ifx\csname bibfnamefont\endcsname\relax
  \def\bibfnamefont#1{#1}\fi
\expandafter\ifx\csname citenamefont\endcsname\relax
  \def\citenamefont#1{#1}\fi
\expandafter\ifx\csname url\endcsname\relax
  \def\url#1{\texttt{#1}}\fi
\expandafter\ifx\csname urlprefix\endcsname\relax\def\urlprefix{URL }\fi
\providecommand{\bibinfo}[2]{#2}
\providecommand{\eprint}[2][]{\url{#2}}	
	

\bibitem[]{LZhang2014}
\bibinfo{author}{\bibfnamefont{L.}~\bibnamefont{Zhang}
and
\bibfnamefont{Q.}~\bibnamefont{Niu}},
  \bibinfo{journal}{Phys. Rev. Lett.} \textbf{\bibinfo{volume}{112}},
  \bibinfo{pages}{085503} (\bibinfo{year}{2014}).

\bibitem[]{LZhang2015}
\bibinfo{author}{\bibfnamefont{L.}~\bibnamefont{Zhang}
and
\bibfnamefont{Q.}~\bibnamefont{Niu}},
  \bibinfo{journal}{Phys. Rev. Lett.} \textbf{\bibinfo{volume}{115}},
  \bibinfo{pages}{115502} (\bibinfo{year}{2015}).  
  

\bibitem[]{Chen2019}
\bibinfo{author}{\bibfnamefont{H.}~\bibnamefont{Chen},
\bibfnamefont{W.}~\bibnamefont{Zhang},
\bibfnamefont{Q.}~\bibnamefont{Niu},
and
\bibfnamefont{L.}~\bibnamefont{Zhang}},
  \bibinfo{journal}{2D Materials} \textbf{\bibinfo{volume}{6}},
  \bibinfo{pages}{012002} (\bibinfo{year}{2019}).    
   
  
\bibitem[]{Rostami2022}
\bibinfo{author}{\bibfnamefont{H.}~\bibnamefont{Rostami},
\bibfnamefont{F.}~\bibnamefont{Guinea},
and
\bibfnamefont{E.}~\bibnamefont{Cappelluti}},
  \bibinfo{journal}{Phys. Rev. B} \textbf{\bibinfo{volume}{105}},
  \bibinfo{pages}{195431} (\bibinfo{year}{2022}).
  
  
\bibitem[]{HZhu2018}
\bibinfo{author}{\bibfnamefont{H.}~\bibnamefont{Zhu},
\bibfnamefont{J.}~\bibnamefont{Yi},
\bibfnamefont{M.-Y.}~\bibnamefont{Li},
\bibfnamefont{J.}~\bibnamefont{Xiao},
\bibfnamefont{L.}~\bibnamefont{Zhang},
\bibfnamefont{C.-W.}~\bibnamefont{Yang},
\bibfnamefont{R.A.}~\bibnamefont{Kaindl},
\bibfnamefont{L.-J.}~\bibnamefont{Li},
\bibfnamefont{Y.}~\bibnamefont{Wang},
and
\bibfnamefont{X.}~\bibnamefont{Zhang}},
  \bibinfo{journal}{Science} \textbf{\bibinfo{volume}{359}},
  \bibinfo{pages}{579} (\bibinfo{year}{2018}).

\bibitem[]{XTChen2019}
\bibinfo{author}{\bibfnamefont{X.}~\bibnamefont{Chen},
\bibfnamefont{X.}~\bibnamefont{Lu},
\bibfnamefont{S.}~\bibnamefont{Dubey},
\bibfnamefont{Q.}~\bibnamefont{Yao},
\bibfnamefont{S.}~\bibnamefont{Liu},
\bibfnamefont{A.}~\bibnamefont{Ataei},
\bibfnamefont{X.}~\bibnamefont{Wang},
\bibfnamefont{M.}~\bibnamefont{Dion},
\bibfnamefont{Q.}~\bibnamefont{Xiong},
\bibfnamefont{L.}~\bibnamefont{Zhang},
and
\bibfnamefont{A.}~\bibnamefont{Srivastava}},
  \bibinfo{journal}{Nat. Phys.} \textbf{\bibinfo{volume}{15}},
  \bibinfo{pages}{221} (\bibinfo{year}{2019}).  

\bibitem[]{Li2019}
\bibinfo{author}{\bibfnamefont{Z.}~\bibnamefont{Li},
\bibfnamefont{T.}~\bibnamefont{Wang},
\bibfnamefont{C.}~\bibnamefont{Jin},
\bibfnamefont{Z.}~\bibnamefont{Lu},
\bibfnamefont{Z.}~\bibnamefont{Lian},
\bibfnamefont{Y.}~\bibnamefont{Meng},
\bibfnamefont{M.}~\bibnamefont{Blei},
\bibfnamefont{M.}~\bibnamefont{Gao},
\bibfnamefont{T.}~\bibnamefont{Taniguchi},
\bibfnamefont{K.}~\bibnamefont{Watanabe},
\bibfnamefont{T.}~\bibnamefont{Ren},
\bibfnamefont{T.}~\bibnamefont{Cao},
\bibfnamefont{S.}~\bibnamefont{Tongay},
\bibfnamefont{D.}~\bibnamefont{Smirnov},
\bibfnamefont{L.}~\bibnamefont{Zhang},
and
\bibfnamefont{S.-F.}~\bibnamefont{Shi}},
  \bibinfo{journal}{ACS Nano} \textbf{\bibinfo{volume}{13}},
  \bibinfo{pages}{14107} (\bibinfo{year}{2019}).

\bibitem[]{Grissonnanche2020}
\bibinfo{author}{\bibfnamefont{G.}~\bibnamefont{Grissonnanche},
\bibfnamefont{S.}~\bibnamefont{Th\'{e}riault},
\bibfnamefont{A.}~\bibnamefont{Gourgout},
\bibfnamefont{M.-E.}~\bibnamefont{Boulanger},
\bibfnamefont{E.}~\bibnamefont{Lefran\c{c}ois},
\bibfnamefont{A.}~\bibnamefont{Ataei},
\bibfnamefont{F.}~\bibnamefont{Lalibert\'{e}},
\bibfnamefont{M.}~\bibnamefont{Dion},
\bibfnamefont{J.-S.}~\bibnamefont{Zhou},
\bibfnamefont{S.}~\bibnamefont{Pyon},
\bibfnamefont{T.}~\bibnamefont{Takayama},
\bibfnamefont{H.}~\bibnamefont{Takagi},
\bibfnamefont{N.}~\bibnamefont{Dorion-Leyraud},
and
\bibfnamefont{T.}~\bibnamefont{Taillefer}},
  \bibinfo{journal}{Nat. Phys.} \textbf{\bibinfo{volume}{16}},
  \bibinfo{pages}{1108} (\bibinfo{year}{2020}). 

\bibitem[]{Jeong2022}
\bibinfo{author}{\bibfnamefont{S.G.}~\bibnamefont{Jeong},
\bibfnamefont{J.}~\bibnamefont{Kim},
\bibfnamefont{A.}~\bibnamefont{Seo},
\bibfnamefont{S.}~\bibnamefont{Park},
\bibfnamefont{H.Y.}~\bibnamefont{Jeong},
\bibfnamefont{Y.-M.}~\bibnamefont{Kim},
\bibfnamefont{V.}~\bibnamefont{Lauter},
\bibfnamefont{T.}~\bibnamefont{Egami},
\bibfnamefont{J.H.}~\bibnamefont{Han},
and
\bibfnamefont{W.S.}~\bibnamefont{Choi}},
  \bibinfo{journal}{Science Advance} \textbf{\bibinfo{volume}{8}},
  \bibinfo{pages}{eabm4005} (\bibinfo{year}{2022}).

\bibitem[]{Ishito2023}
\bibinfo{author}{\bibfnamefont{K.}~\bibnamefont{Ishito},
\bibfnamefont{H.}~\bibnamefont{Mao},
\bibfnamefont{Y.}~\bibnamefont{Kousaka},
\bibfnamefont{Y.}~\bibnamefont{Togawa},
\bibfnamefont{S.}~\bibnamefont{Iwasaki},
\bibfnamefont{T.}~\bibnamefont{Zhang},
\bibfnamefont{S.}~\bibnamefont{Murakami},
\bibfnamefont{J.-I.}~\bibnamefont{Kishine},
and
\bibfnamefont{T.}~\bibnamefont{Satoh}},
  \bibinfo{journal}{Nat. Phys.} \textbf{\bibinfo{volume}{19}},
  \bibinfo{pages}{35} (\bibinfo{year}{2023}).  

\bibitem[]{Ueda2023}
\bibinfo{author}{\bibfnamefont{H.}~\bibnamefont{Ueda},
\bibfnamefont{M.}~\bibnamefont{Garc\'{i}a-Fern\'{a}ndez},
\bibfnamefont{S.}~\bibnamefont{Agrestini},
\bibfnamefont{C.}~\bibnamefont{P.Romao},
\bibfnamefont{J.}~\bibnamefont{van den Brink},
\bibfnamefont{N.A.}~\bibnamefont{Spaldin},
\bibfnamefont{K.-J.}~\bibnamefont{Zhou},
and
\bibfnamefont{U.}~\bibnamefont{Staub}},
  \bibinfo{journal}{Nature} \textbf{\bibinfo{volume}{618}},
  \bibinfo{pages}{946} (\bibinfo{year}{2023}).

  

\bibitem[]{Hamada2018}
\bibinfo{author}{\bibfnamefont{M.}~\bibnamefont{Hamada},
\bibfnamefont{E.}~\bibnamefont{Minamitani},
\bibfnamefont{M.}~\bibnamefont{Hirayama},
and
\bibfnamefont{S.}~\bibnamefont{Murakami}},
  \bibinfo{journal}{Phys. Rev. Lett.} \textbf{\bibinfo{volume}{121}},
  \bibinfo{pages}{175301} (\bibinfo{year}{2018}).

\bibitem[]{Hamada2020}
\bibinfo{author}{\bibfnamefont{M.}~\bibnamefont{Hamada}
and
\bibfnamefont{S.}~\bibnamefont{Murakami}},
  \bibinfo{journal}{Phys. Rev. B} \textbf{\bibinfo{volume}{101}},
  \bibinfo{pages}{144306} (\bibinfo{year}{2020}).


\bibitem[]{Nova2017}
\bibinfo{author}{\bibfnamefont{T. F.}~\bibnamefont{Nova},
\bibfnamefont{A.}~\bibnamefont{Cartella},
\bibfnamefont{A.}~\bibnamefont{Cantaluppi},
\bibfnamefont{M.}~\bibnamefont{F\"{o}rst},
\bibfnamefont{D.}~\bibnamefont{Bossini},
\bibfnamefont{R. V.}~\bibnamefont{Mikhaylovskiy},
\bibfnamefont{A. V.}~\bibnamefont{Kimel},
\bibfnamefont{R.}~\bibnamefont{Merlin},
and
\bibfnamefont{A.}~\bibnamefont{Cavalleri}},
  \bibinfo{journal}{Nat. Phys.} \textbf{\bibinfo{volume}{13}},
  \bibinfo{pages}{132} (\bibinfo{year}{2017}). 

\bibitem[]{Juraschek2019}
\bibinfo{author}{\bibfnamefont{D.M.}~\bibnamefont{Juraschek}
and
\bibfnamefont{N.A.}~\bibnamefont{Spaldin}},
  \bibinfo{journal}{Phys. Rev. Mater.} \textbf{\bibinfo{volume}{13}},
  \bibinfo{pages}{132} (\bibinfo{year}{2019}).
  
\bibitem[]{Geilhufe2021}
\bibinfo{author}{\bibfnamefont{R.M.}~\bibnamefont{Geilhufe},
\bibfnamefont{V.}~\bibnamefont{Juri\v{c}i\'{c}},
\bibfnamefont{S.}~\bibnamefont{Bonetti},
\bibfnamefont{J.-X.}~\bibnamefont{Zhu},
and
\bibfnamefont{A.V.}~\bibnamefont{Balatsky}},
  \bibinfo{journal}{Phys. Rev. Res.} \textbf{\bibinfo{volume}{3}},
  \bibinfo{pages}{144302} (\bibinfo{year}{2022}).

\bibitem[]{Juraschek2022}
\bibinfo{author}{\bibfnamefont{D.M.}~\bibnamefont{Juraschek},
\bibfnamefont{T.}~\bibnamefont{Neuman},
and
\bibfnamefont{P.}~\bibnamefont{Narang}},
  \bibinfo{journal}{Phys. Rev. Res.} \textbf{\bibinfo{volume}{4}},
  \bibinfo{pages}{013129} (\bibinfo{year}{2022}).

\bibitem[]{Xiong2022}
\bibinfo{author}{\bibfnamefont{G.}~\bibnamefont{Xiong},
\bibfnamefont{H.}~\bibnamefont{Chen},
\bibfnamefont{D.}~\bibnamefont{Ma},
and
\bibfnamefont{L.}~\bibnamefont{Zhang}},
  \bibinfo{journal}{Phys. Rev. B} \textbf{\bibinfo{volume}{106}},
  \bibinfo{pages}{144302} (\bibinfo{year}{2022}).

\bibitem[]{Luo2023}
\bibinfo{author}{\bibfnamefont{J.}~\bibnamefont{Luo},
\bibfnamefont{T.}~\bibnamefont{Lin},
\bibfnamefont{J.}~\bibnamefont{Zhang},
\bibfnamefont{X.}~\bibnamefont{Chen},
\bibfnamefont{E.R.}~\bibnamefont{Blackert},
\bibfnamefont{R.}~\bibnamefont{Xu},
\bibfnamefont{B.I.}~\bibnamefont{Yakubson},
and
\bibfnamefont{H.}~\bibnamefont{Zhu}},
  \bibinfo{journal}{Science} \textbf{\bibinfo{volume}{382}},
  \bibinfo{pages}{698} (\bibinfo{year}{2023}).

\bibitem[]{Hernandez2022}
\bibinfo{author}{\bibfnamefont{F.G.G.}~\bibnamefont{Hernandez},
\bibfnamefont{A.}~\bibnamefont{Baydin},
\bibfnamefont{S.}~\bibnamefont{Chaudhary},
\bibfnamefont{F.}~\bibnamefont{Tay},
\bibfnamefont{I.}~\bibnamefont{Katayama},
\bibfnamefont{J.}~\bibnamefont{Takeda},
\bibfnamefont{H.}~\bibnamefont{Nojiri},
\bibfnamefont{A.K.}~\bibnamefont{Okazaki},
\bibfnamefont{P.H.O.}~\bibnamefont{Rappl},
\bibfnamefont{E.}~\bibnamefont{Abramof},
\bibfnamefont{M.}~\bibnamefont{Rodriguez-Vega},
\bibfnamefont{G.A.}~\bibnamefont{Fiete},
and
\bibfnamefont{J.}~\bibnamefont{Kono},
  \bibinfo{journal}{Science Advance} 
  \textbf{\bibinfo{volume}{9}},
  \bibinfo{pages}{eadj4074}}
   (\bibinfo{year}{2023}).
      
\bibitem[]{Chaudhary2023}
\bibinfo{author}{\bibfnamefont{S.}~\bibnamefont{Chaudhary},
\bibfnamefont{D.M.}~\bibnamefont{Juraschek},
\bibfnamefont{M.}~\bibnamefont{Rodriguez-Vega},
and
\bibfnamefont{G.A.}~\bibnamefont{Fiete},
  \bibinfo{journal}{arXiv:2306.11630}}
   (\bibinfo{year}{2023}).
 

\bibitem[]{HamadaPRR2020}
\bibinfo{author}{\bibfnamefont{M.}~\bibnamefont{Hamada}
and
\bibfnamefont{S.}~\bibnamefont{Murakami}},
  \bibinfo{journal}{Phys. Rev. Res.} \textbf{\bibinfo{volume}{2}},
  \bibinfo{pages}{023275} (\bibinfo{year}{2020}).

\bibitem[]{Ren2021}
\bibinfo{author}{\bibfnamefont{Y.}~\bibnamefont{Ren},
\bibfnamefont{C.}~\bibnamefont{Xiao},
\bibfnamefont{D.}~\bibnamefont{Saparov},
and
\bibfnamefont{Q.}~\bibnamefont{Niu}},
  \bibinfo{journal}{Phys. Rev. Lett.} \textbf{\bibinfo{volume}{127}},
  \bibinfo{pages}{186403} (\bibinfo{year}{2021}).

\bibitem[]{Yao2022}
\bibinfo{author}{\bibfnamefont{D.}~\bibnamefont{Yao}
and
\bibfnamefont{S.}~\bibnamefont{Murakami}},
  \bibinfo{journal}{Phys. Rev. B} \textbf{\bibinfo{volume}{105}},
  \bibinfo{pages}{184412} (\bibinfo{year}{2022}).

\bibitem[]{Tauchert2022}
\bibinfo{author}{\bibfnamefont{S.R.}~\bibnamefont{Tauchert},
\bibfnamefont{M.}~\bibnamefont{Volkov},
\bibfnamefont{D.}~\bibnamefont{Ehberger},
\bibfnamefont{D.}~\bibnamefont{Kazenwadel},
\bibfnamefont{M.}~\bibnamefont{Evers},
\bibfnamefont{H.}~\bibnamefont{Lange},
\bibfnamefont{A.}~\bibnamefont{Donges},
\bibfnamefont{A.}~\bibnamefont{Book},
\bibfnamefont{W.}~\bibnamefont{Kreuzpaintner},
\bibfnamefont{U.}~\bibnamefont{Nowak},
and
\bibfnamefont{P.}~\bibnamefont{Baum}},
  \bibinfo{journal}{Nature} \textbf{\bibinfo{volume}{602}},
  \bibinfo{pages}{73} (\bibinfo{year}{2022}).

\bibitem[]{Fransson2023}
\bibinfo{author}{\bibfnamefont{J.}~\bibnamefont{Fransson}},
  \bibinfo{journal}{Phys. Rev. Res.} \textbf{\bibinfo{volume}{5}},
  \bibinfo{pages}{L022039} (\bibinfo{year}{2023}). 

\bibitem[]{Yao2023}
\bibinfo{author}{\bibfnamefont{D.}~\bibnamefont{Yao}
and
\bibfnamefont{S.}~\bibnamefont{Murakami}},
  \bibinfo{journal}{J. Phys. Soc. Jpn.} 
  \textbf{\bibinfo{volume}{93}},
  \bibinfo{pages}{034708}
   (\bibinfo{year}{2024}).

\bibitem[]{Davies2023}
\bibinfo{author}{\bibfnamefont{C.S.}~\bibnamefont{Davies},
\bibfnamefont{F.G.N.}~\bibnamefont{Fennema},
\bibfnamefont{A.}~\bibnamefont{Tsukamoto},
\bibfnamefont{I.}~\bibnamefont{Razdolski},
\bibfnamefont{A.V.}~\bibnamefont{Kimel},
and
\bibfnamefont{A.}~\bibnamefont{Kirilyuk},
  \bibinfo{journal}{arXiv:2305.11551}} 
   (\bibinfo{year}{2023}).

\bibitem[]{YafenRen2023}
\bibinfo{author}{\bibfnamefont{Y.}~\bibnamefont{Ren},
\bibfnamefont{M.}~\bibnamefont{Rudner},
and
\bibfnamefont{D.}~\bibnamefont{Xiao}},
  \bibinfo{journal}{arXiv:2308.00933}
   (\bibinfo{year}{2023}).
   
\bibitem[]{Kahana2023}
\bibinfo{author}{\bibfnamefont{T.}~\bibnamefont{Kahana},
\bibfnamefont{D.A.B}~\bibnamefont{Lopez},
and
\bibfnamefont{D.M.}~\bibnamefont{Juraschek}},
  \bibinfo{journal}{arXiv:2305.18656}
   (\bibinfo{year}{2023}).   


\bibitem[]{Kim2023}
\bibinfo{author}{\bibfnamefont{K.}~\bibnamefont{Kim},
\bibfnamefont{E.}~\bibnamefont{Vetter},
\bibfnamefont{L.}~\bibnamefont{Yan},
\bibfnamefont{C.}~\bibnamefont{Yang},
\bibfnamefont{Z.}~\bibnamefont{Wang},
\bibfnamefont{R.}~\bibnamefont{Sun},
\bibfnamefont{Y.}~\bibnamefont{Yang},
\bibfnamefont{A.H.}~\bibnamefont{Comstock},
\bibfnamefont{X.}~\bibnamefont{Li},
\bibfnamefont{J.}~\bibnamefont{Zhou},
\bibfnamefont{L.}~\bibnamefont{Zhang},
\bibfnamefont{W.}~\bibnamefont{You},
\bibfnamefont{D.}~\bibnamefont{Sun},
and
\bibfnamefont{J.}~\bibnamefont{Liu}},
  \bibinfo{journal}{Nature Materials} \textbf{\bibinfo{volume}{22}},
  \bibinfo{pages}{322} (\bibinfo{year}{2023}).
  
\bibitem[]{Li2022}
\bibinfo{author}{\bibfnamefont{X.}~\bibnamefont{Li},
\bibfnamefont{J.}~\bibnamefont{Zhong},
\bibfnamefont{J.}~\bibnamefont{Cheng},
\bibfnamefont{H.}~\bibnamefont{Chen},
\bibfnamefont{H.}~\bibnamefont{Wang},
\bibfnamefont{J.}~\bibnamefont{Liu},
\bibfnamefont{D.}~\bibnamefont{Sun},
\bibfnamefont{L.}~\bibnamefont{Zhang},
and
\bibfnamefont{J.}~\bibnamefont{Zhou}},
  \bibinfo{journal}{Sci.~China-Phys.~Mech.~Astron.} 
  \textbf{\bibinfo{volume}{67}},
  \bibinfo{pages}{237511}
   (\bibinfo{year}{2024}).


\bibitem[]{Nakajima2023}
\bibinfo{author}{\bibfnamefont{R.}~\bibnamefont{Nakajima},
\bibfnamefont{D.}~\bibnamefont{Hirobe},
\bibfnamefont{G.}~\bibnamefont{Kawaguchi},
\bibfnamefont{Y.}~\bibnamefont{Nabei},
\bibfnamefont{T.}~\bibnamefont{Sato},
\bibfnamefont{T.}~\bibnamefont{Narushima},
\bibfnamefont{H.}~\bibnamefont{Okamoto},
and
\bibfnamefont{H. M.}~\bibnamefont{Yamamoto}},
  \bibinfo{journal}{Nature} \textbf{\bibinfo{volume}{613}},
  \bibinfo{pages}{479} (\bibinfo{year}{2023}).


\bibitem[]{Perrin1974}
\bibinfo{author}{\bibfnamefont{N.}~\bibnamefont{Perrin}
and
\bibfnamefont{H.}~\bibnamefont{Budd}}
  \bibinfo{journal}{Phys. Rev. B} \textbf{\bibinfo{volume}{9}},
  \bibinfo{pages}{3454} (\bibinfo{year}{1974}).

\bibitem[]{Gurevich1989}
\bibinfo{author}{\bibfnamefont{Yu.G.}~\bibnamefont{Gurevich}
and
\bibfnamefont{O.L.}~\bibnamefont{Mashkevich}}
  \bibinfo{journal}{Physics Reports} \textbf{\bibinfo{volume}{181}},
  \bibinfo{pages}{327} (\bibinfo{year}{1989}).

\bibitem[]{Lozel1975}
\bibinfo{author}{\bibfnamefont{V.A.}~\bibnamefont{Kizel'},
\bibfnamefont{Y.}~\bibnamefont{Krasilov},
and
\bibfnamefont{V.I.}~\bibnamefont{Burkov}},
  \bibinfo{journal}{Sov. Phys. Usp.} \textbf{\bibinfo{volume}{17}},
  \bibinfo{pages}{745} (\bibinfo{year}{1975}).

\bibitem[]{Jerphagnon1976}
\bibinfo{author}{\bibfnamefont{J.}~\bibnamefont{Jerphagnon}
and
\bibfnamefont{D.S.}~\bibnamefont{Chemla}},
  \bibinfo{journal}{J. Chem. Phys.} \textbf{\bibinfo{volume}{65}},
  \bibinfo{pages}{1522} (\bibinfo{year}{1976}).

\bibitem[]{Ganichev2016}
\bibinfo{author}{\bibfnamefont{S.D.}~\bibnamefont{Ganichev},
\bibfnamefont{M.}~\bibnamefont{Trushin},
and
\bibfnamefont{J.}~\bibnamefont{Schliemann}},
  \bibinfo{journal}{arXiv:1606.02043} 
   (\bibinfo{year}{2016}).


\bibitem[]{Edelstein1990}
\bibinfo{author}{\bibfnamefont{V.M.}~\bibnamefont{Edelstein}},
  \bibinfo{journal}{Solid State Communications} \textbf{\bibinfo{volume}{73}},
  \bibinfo{pages}{233} (\bibinfo{year}{1990}).
  
\bibitem[]{Edelstein1995}
\bibinfo{author}{\bibfnamefont{V.M.}~\bibnamefont{Edelstein}},
  \bibinfo{journal}{Phys. Rev. Lett.} \textbf{\bibinfo{volume}{75}},
  \bibinfo{pages}{2004} (\bibinfo{year}{1995}).

\bibitem[]{Edelstein2005}
\bibinfo{author}{\bibfnamefont{V.M.}~\bibnamefont{Edelstein}},
  \bibinfo{journal}{Phys. Rev. B} \textbf{\bibinfo{volume}{72}},
  \bibinfo{pages}{172501} (\bibinfo{year}{2005}).  
  
\bibitem[]{He}
\bibinfo{author}{\bibfnamefont{W.-Y.}~\bibnamefont{He}}, and
\bibfnamefont{K.T.}~\bibnamefont{Law},
  \bibinfo{journal}{Phys. Rev. Res.} \textbf{\bibinfo{volume}{2}},
  \bibinfo{pages}{012073} (\bibinfo{year}{2020}).  
  

\bibitem[]{Tinkham}
\bibinfo{author}{\bibfnamefont{M.}~\bibnamefont{Tinkham}},
  \emph{\bibinfo{title}{Introduction to Superconductivity}}
  (\bibinfo{publisher}{Dover Publications},
  \bibinfo{address}{New Yark}, \bibinfo{year}{2004}), \bibinfo{edition}{2nd}
  edition.


\bibitem[]{Yamashita2002}
\bibinfo{author}{\bibfnamefont{T.}~\bibnamefont{Yamashita},
\bibfnamefont{S.}~\bibnamefont{Takahashi},
\bibfnamefont{H.}~\bibnamefont{Imamura},
and
\bibfnamefont{S.}~\bibnamefont{Maekawa}},
  \bibinfo{journal}{Phys. Rev. B} \textbf{\bibinfo{volume}{65}},
  \bibinfo{pages}{172509} (\bibinfo{year}{2002}).
 
\bibitem[]{Gao2023}
\bibinfo{author}{\bibfnamefont{Y.}~\bibnamefont{Gao},
\bibfnamefont{Y.}~\bibnamefont{Pan},
\bibfnamefont{J.}~\bibnamefont{Zhou},
and
\bibfnamefont{L.}~\bibnamefont{Zhang}},
  \bibinfo{journal}{Phys. Rev. B} \textbf{\bibinfo{volume}{108}},
  \bibinfo{pages}{064510} (\bibinfo{year}{2023}).
 
  
\bibitem[]{Amoretti2022}
\bibinfo{author}{\bibfnamefont{A.}~\bibnamefont{Amoretti},
\bibfnamefont{D.K.}~\bibnamefont{Brattan},
\bibfnamefont{N.}~\bibnamefont{Magnoli},
\bibfnamefont{L.}~\bibnamefont{Martinoia},
\bibfnamefont{I.}~\bibnamefont{Matthaiakakis},
and
\bibfnamefont{P.}~\bibnamefont{Solinas}},
  \bibinfo{journal}{Phys. Rev. Res.} \textbf{\bibinfo{volume}{4}},
  \bibinfo{pages}{033211} (\bibinfo{year}{2022}).
  
  
\bibitem[]{Ando2011}
\bibinfo{author}{\bibfnamefont{K.}~\bibnamefont{Ando},
\bibfnamefont{S.}~\bibnamefont{Takahashi},
\bibfnamefont{J.}~\bibnamefont{Ieda},
\bibfnamefont{Y.}~\bibnamefont{Kajiwara},
\bibfnamefont{H.}~\bibnamefont{Nakayama},
\bibfnamefont{T.}~\bibnamefont{Yoshino},
\bibfnamefont{K.}~\bibnamefont{Harii},
\bibfnamefont{Y.}~\bibnamefont{Fujikawa},
\bibfnamefont{M.}~\bibnamefont{Matsuo},
\bibfnamefont{S.}~\bibnamefont{Maekawa},
and
\bibfnamefont{E.}~\bibnamefont{Saitoh}},
  \bibinfo{journal}{J. Appl. Phys.} \textbf{\bibinfo{volume}{109}},
  \bibinfo{pages}{103913} (\bibinfo{year}{2011}).
  
\bibitem[]{Takahashi2008}
\bibinfo{author}{\bibfnamefont{S.}~\bibnamefont{Takahashi}
and
\bibfnamefont{S.}~\bibnamefont{Maekawa}},
  \bibinfo{journal}{Sci. Technol. Adv. Mater.} \textbf{\bibinfo{volume}{9}},
  \bibinfo{pages}{014105} (\bibinfo{year}{2008}).  

\bibitem[]{Tateno2020}
\bibinfo{author}{\bibfnamefont{S.}~\bibnamefont{Tateno},
\bibfnamefont{G.}~\bibnamefont{Okano},
\bibfnamefont{M.}~\bibnamefont{Matsuo},
and
\bibfnamefont{Y.}~\bibnamefont{Nozaki}},
  \bibinfo{journal}{Phys. Rev. B} \textbf{\bibinfo{volume}{102}},
  \bibinfo{pages}{104406} (\bibinfo{year}{2020}).  
  
  
\end{thebibliography}
\end{document}